\documentclass[runningheads]{llncs}
\usepackage[T1]{fontenc}
\usepackage{comment}
\usepackage{amsmath,amssymb}
\usepackage{color}
\PassOptionsToPackage{hyphens}{url}\usepackage{hyperref}
\usepackage{hyperref}
\usepackage{graphicx,verbatim}
\usepackage{booktabs}
\usepackage{multirow} 
\begin{document}
\title{Enhancing Synthetic CT from CBCT via Multimodal Fusion and End-To-End Registration}
\titlerunning{Enhancing Synthetic CT via Multimodal Fusion and Registration}
%
%\titlerunning{Abbreviated paper title}
% If the paper title is too long for the running head, you can set
% an abbreviated paper title here
%

%\author{Anonymized Authors}  %% Added for anonymized MICCAI 2025 submission
%\authorrunning{Anonymized Author et al.}
%\institute{Anonymized Affiliations \\
%    \email{email@anonymized.com}}

\author{Maximilian Tschuchnig\inst{1,3}\orcidID{0000-0002-1441-4752} \and
Lukas Lamminger\inst{2} \and
Philipp Steininger\inst{2} \and
Michael Gadermayr\inst{1}\orcidID{0000-0003-1450-9222}}

\authorrunning{M. Tschuchnig et al.}

\institute{Salzburg University of Applied Sciences, Puch bei Hallein, Austria \and MedPhoton GmbH, Salzburg, Austria \and University of Salzburg, Salzburg, Austria
\email{maximilian.tschuchnig@fh-salzburg.ac.at}}

\maketitle              % typeset the header of the contribution
\begin{abstract}
Cone-Beam Computed Tomography (CBCT) is widely used for intraoperative imaging due to its rapid acquisition and low radiation dose. However, CBCT images typically suffer from artifacts and lower visual quality compared to conventional Computed Tomography (CT). A promising solution is synthetic CT (sCT) generation, where CBCT volumes are translated into the CT domain. In this work, we enhance sCT generation through multimodal learning by jointly leveraging intraoperative CBCT and preoperative CT data. To overcome the inherent misalignment between modalities, we introduce an end-to-end learnable registration module within the sCT pipeline. This model is evaluated on a controlled synthetic dataset, allowing precise manipulation of data quality and alignment parameters. Further, we validate its robustness and generalizability on two real-world clinical datasets. Experimental results demonstrate that integrating registration in multimodal sCT generation improves sCT quality, outperforming baseline multimodal methods in $79$ out of $90$ evaluation settings. Notably, the improvement is most significant in cases where CBCT quality is low and the preoperative CT is moderately misaligned.

\end{abstract}
\section{Introduction}
\sloppy Mobile robotic imaging systems, such as cone-beam computed tomography (CBCT)~\cite{rafferty2006intraoperative} provide real-time, intraoperative imaging, facilitating guidance during medical procedures. This is especially useful, for example, in radiation therapy~\cite{thummerer2023synthrad}. CBCT uses a cone-shaped X-ray beam to acquire 3D images in a single rotation, reducing acquisition time and radiation exposure. However, CBCT images typically suffer from lower contrast-to-noise ratios and increased artifacts compared to CT scans~\cite{wei2024reduction}.

One approach to mitigating CBCT artifacts, as highlighted by Altalib et al.~\cite{altalib2025synthetic}, is the conversion of high-artifact CBCT volumes into a low-artifact CT domain using image-to-image translation. sCTs~\cite{altalib2025synthetic,chen2020synthetic} aim to leverage domain advantages, here the acquisition speed and mobility of CBCT while improving image quality by reducing artifacts. The importance of robust sCT was underscored by the SynthRad Grand Challenge 2023~\cite{thummerer2023synthrad}, which exposed key limitations of existing methods, especially their sensitivity to CBCT artifacts and anatomical misalignment, and motivated the need for techniques that are more robust to artifacts and anatomical misalignment.

However, sCT generation faces two key limitations. The methods cannot create new anatomical details beyond those present in the input data, and the effectiveness of training models generating sCT depends on the availability of high-quality training data. While this second limitation can be partially addressed by data augmentation, addressing the first limitation requires integrating additional information such as high-quality preoperative CT scans into the sCT generation process~\cite{tschuchnig2025enhancing,altalib2025synthetic,chen2021synthetic,chen2020synthetic}. Tschuchnig et al.~\cite{tschuchnig2025enhancing}, for example, combine high quality, preoperativce CT with intraoperative CBCT for multimodal sCT generation. These high-quality, preoperative CT scans are typically acquired before a procedure to aid in treatment planning. For instance, in radiotherapy, treatment plans are initially based on high-resolution preoperative CT scans~\cite{hong2022ct}. However, since intraoperative CBCT and preoperative CT are typically acquired at different times and under varying anatomical conditions, accurately aligning them remains an essential component of any multimodal sCT pipeline.

As shown by Tschuchnig et al., the fusion of CBCT-CT for multimodal sCT improves on unimodal sCT approaches in most cases~\cite{tschuchnig2025enhancing,zhao2024emma}. This fusion allows the model to learn from the complementary strengths of both modalities: CBCT provides real-time anatomical information during the procedure, while CT offers high-quality structural detail acquired during planning. Multimodal learning, effectively leverages this complementary information. %Recent advances in multi-modality fusion from computer vision, such as the self-supervised Equivariant Multi-Modality Image Fusion framework for heterogeneous data integration~\cite{zhao2024emma}, further demonstrate the potential to effectively combine multiple imaging sources. 
However, a key challenge in multimodal sCT generation is the inherent misalignment between intraoperative CBCT and preoperative CT, caused by anatomical changes and differences in patient positioning. While traditional multimodal fusion approaches often rely on external pre-registration, Tschuchnig et al.~\cite{tschuchnig2025enhancing} show that U-Net based multimodal approaches improve on unimodal, even in misaligned scenarios. However, they still reported a strong gap between their observed implicit registration and potential additional registration. We focus on integrating a learnable registration component directly into the multimodal sCT generation pipeline. Specifically, we employ Spatial Transformer Networks (STN)~\cite{jaderberg2015spatial} to enable end-to-end optimization of both the registration and synthesis tasks, allowing the network to jointly optimize registration and synthesis, and adaptively align preoperative CT to intraoperative CBCT during training.

The main contributions of this work are manifold. First, we propose an end-to-end multimodal sCT generation framework that integrates a STN to explicitly address anatomical misalignment between intraoperative CBCT and preoperative CT. Second, we conduct an extensive evaluation on the effect of the STN component, comparing performance against conventional unimodal and multimodal baselines without registration. Third, using a synthetic dataset with controlable quality parameters, we investigate the effects of CBCT quality and CBCT-CT alignment on the effect of STN. Fourth, we demonstrate the robustness and reproducibility of our approach across two real-world clinical datasets.

\section{Methodology and Materials}
\begin{figure}
\includegraphics[width=\textwidth]{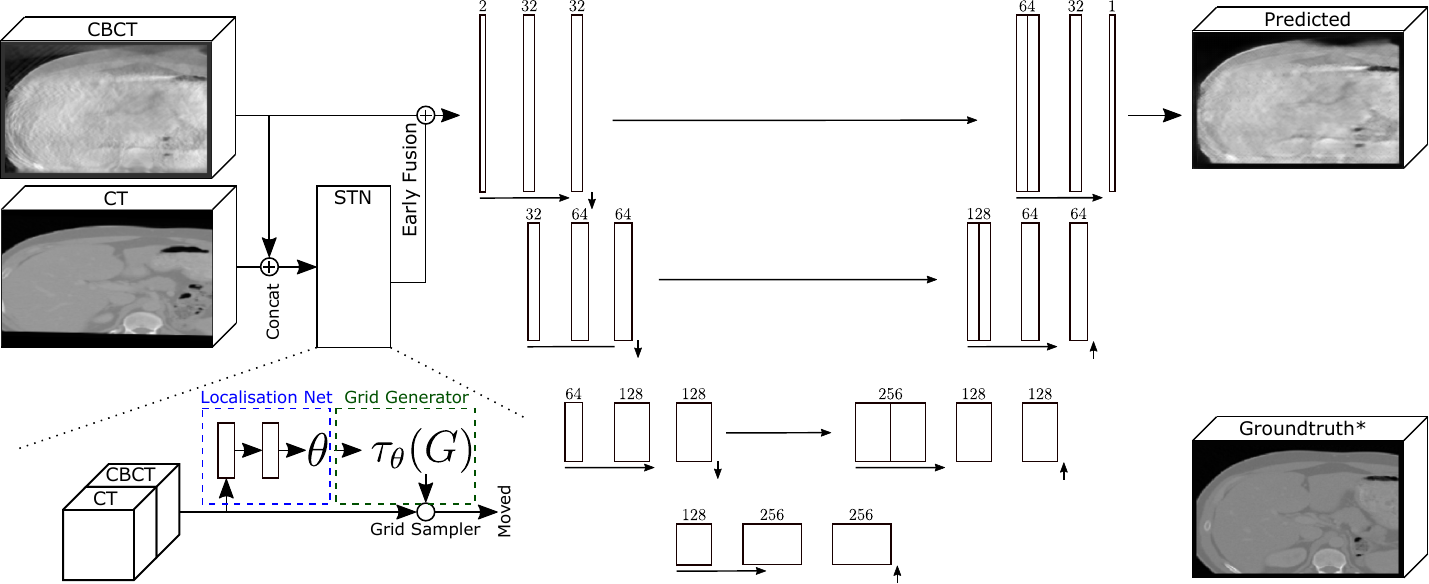}
\caption{Proposed 3D, (early fusion) multimodal sCT model, based on U-Net.} \label{Fig::I2IMethod}
\end{figure}

The proposed 3D multimodal sCT generation method combines deep learning with multimodal learning by adapting a 3D image reconstruction model, specifically 3D U-Net~\cite{cciccek20163d}, with early fusion multimodal learning~\cite{podobnik2023multimodal}. 
To facilitate CBCT-CT alignment, a STN module is added before the first layer of the 3D U-Net, aligning the preoperative CT with the intraoperative CBCT. The full architecture is illustrated in Fig.~\ref{Fig::I2IMethod}.
The model $F$ takes as input a combination of preoperative, unaligned CT volumes $U_{CT}$ and intraoperative CBCT volumes $V_{CBCT}$. $U_{CT}$ is aligned to $V_{CBCT}$ using the STN. To accomplish this the STN takes the concatenation of $[U_{CT}, V_{CBCT}]$ to estimate affine registration parameters, aimed at registering $U_{CT}$ to $V_{CBCT}$, resulting in $V_{CT}$. Then, the volumes $V_{CBCT}$ and $V_{CT}$ are concatenated along a fourth dimension, analogous to how RGB channels are treated in 2D images, to form the combined volume $V=[V_{CT},V_{CBCT}]$. This volume is then fed into a reconstruction U-Net to generate the corresponding sCT $\hat{Y}$. Using $\hat{Y}$ and the preoperative, aligned CT, $Y$, the loss is calculated with $L(\hat{Y},Y)$. 

Specifically, the model begins with an STN component, followed by an encoder consisting of three double convolutional blocks, each with $3 \times 3 \times 3$ kernels, connected by 3D max pooling layers. The latent space contains one additional double convolutional block, followed by a decoder that mirrors the encoder structure. Each encoder block is connected to its corresponding decoder block via skip connections. To facilitate final image reconstruction, a $1 \times 1 \times 1$ 3D convolutional layer with a single output channel is added at the end of the decoder. Since no transfer function is applied to this layer, it outputs logits directly. The number of feature maps at each level is $\{32, 64, 128, 256\}$, and batch normalization is applied after each convolutional layer.

The STN module is implemented as a CNN with three 3D convolutional layers containing $\{16, 32, 64\}$ filters and kernel sizes of $\{7, 5, 3\}$, respectively. Max pooling is applied after the first convolutional layer. Following the convolutional stack, adaptive average pooling is used to compress the spatial dimensions before passing the features to the parameter estimation head. This head consists of two fully connected layers with output sizes $\{32, 12\}$. The $12$ predicted parameters represent an affine transformation, which is then converted into a sampling grid and applied to the moving (preoperative) CT using a grid sampler. Both the U-Net and STN use ReLU activation functions throughout.

To train the model we use a weighted sum of losses. Specifically, voxel based, patch based and perceptual losses, as typically used in image reconstruction and image-to-image based models~\cite{johnson2016perceptual,chen2020synthetic,altalib2025synthetic}. As a voxel based loss we apply mean absolute error (MAE) to focus on preserving image structure and robustness to noise. As a patch based loss, the structured similarity index measure (SSIM) is applied to also focus on the patch information luminance, contrast and structure. To convert the similarity into a loss function, we utilize $1-SSIM(\hat{Y},Y)$.
Additionally, a perceptual loss~\cite{johnson2016perceptual} based on vgg16~\cite{Simonyan2014VeryDC}, pretrained on the imagenet dataset~\cite{deng2009imagenet} is added to preserve perceptual information. In detail, we use the pretrained vgg16 to extract mid-level feature maps (vgg16 up to and including layer $23$) from the sCT and compare these feature maps to the original CTs vgg16 feature maps using MAE. Therefore, our perceptual loss is defined as $MAE(vgg_{:23}(\hat{Y}),vgg_{:23}(Y))$. The final U-Net loss is given by $\alpha_1 \cdot MAE(\hat{Y},Y) + \alpha_2 \cdot 1-SSIM(\hat{Y},Y) + \alpha_3 \cdot MAE(vgg_{:23}(\hat{Y}),vgg_{:23}(Y))$ with $\alpha_1=0.2$, $\alpha_2=0.1$ and $\alpha_3=0.7$. To tune the STN, an additional registration loss is added, defined as $10^{-3} \cdot MSE(V_{CT},Y)$.

Loss weights were empirically chosen based on validation metrics and qualitative assessment. The perceptual loss was prioritized to improve sharpness and anatomical detail. In qualitative observations, relying solely on perceptual loss led to hallucinated structures, which were mitigated by incorporating the voxel-wise MAE term. SSIM further complemented both by encouraging local structural coherence and contrast preservation. To facilitate reproducibility, all source code used for these experiments is publicly available on GitHub at: \url{https://github.com/MaxTschuchnig/EnhancingSyntheticCTfromCBCTviaMultimodalFusionandEnd-To-EndRegistration}.

As baselines, unimodal sCT models are applied using only CBCT as input and multimodal models without STN~\cite{tschuchnig2025enhancing}. To assess whether the multimodal model primarily relies on the unaligned CT while disregarding the intraoperative CBCT, we introduce an additional baseline (CT-only), which uses only the unaligned CT as input and compares the output to the perfectly aligned CT. However, due to the inherent unalignment of CBCT and CT, this baseline is imperfect in real-world datasets.

\subsection{Dataset}
To investigate the effect of the STN, three datasets, one synthetic and two real-world datasets are used:

\begin{enumerate}
  \item CBCT Liver Tumor Segmentation Benchmark (CBCTLiTS) dataset~\cite{tschuchnig2024cbctlits}. CBCTLiTS consists of $131$ synthetic CBCT and corresponding real CT images of the abdominal region. The paired volumes are perfectly aligned and available in five different quality levels, enabling a controlled study of intraoperative image quality ($\alpha_{np}$). 
  \item Pancreatic-CT-CBCT-SEG dataset~\cite{hong2022ct} (pancreas dataset) The pancreas dataset provides $40$ CBCT-CT pairs of the abdominal region, serving as a challenging benchmark to validate our findings in a clinical setting. Unlike CBCTLiTS, the pancreas dataset includes naturally occurring misalignment and relatively uniform CBCT quality.  
  \item SynthRAD2023~\cite{thummerer2023synthrad} dataset (synthrad). Synthrad provides 360 paired CT-CBCT and CT-MR volumes across brain and pelvis anatomies, acquired from three clinical centers. The dataset includes both rigidly registered CBCT and MRI images to planning CTs, enabling sCT generation tasks across varying acquisition protocols and image qualities. For our experiments, we chose the target of pelvis sCT generation from CBCT.
\end{enumerate}

All datasets used are publicly available: CBCTLiTS under CC BY-NC-ND 4.0, Pancreatic-CT-CBCT-SEG under CC BY 4.0, and SynthRAD under CC BY-NC 4.0.

\begin{figure}[t]
\includegraphics[width=\textwidth]{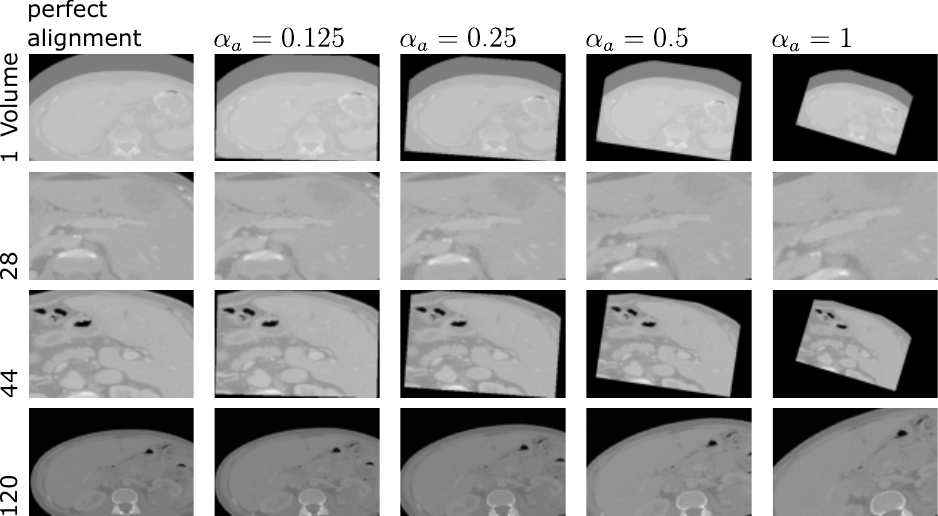}
\caption{Sample results, showing the original CT as well as synthetically unaligned version of the same CT (CBCTLiTS dataset). Four different volumes are shown with increasing $\alpha_a$ (unalignment), displaying rotation, scaling and minor translation.} \label{Fig::DataAugResults}
\end{figure}

In the real-world datasets (pancreas and synthrad), CBCT-CT alignment is not perfect (even if performed manually or semi-automatically) due to patient movement and respiratory variations. Even with controlled breathing techniques designed to minimize motion-induced misalignment, perfect CBCT-CT alignment remains difficult to achieve~\cite{hong2022ct}. This is an important detail which potentially affects both, training (training with imperfect pairs) and evaluation (applying measures on imperfectly aligned pairs).

To assess the impact of alignment on multimodal sCT reconstruction, the parameter $\alpha_a$ is introduced. This factor, defined as $\alpha_a$, describes how strongly the preoperative CT is artificially unaligned. $\alpha_a$ reduces the number of parameters controlling unalignment by combining multiple affine unalignments (rotation, scaling and translation) into a single parameter. Artificial unalignment was performed using TorchIO RandomAffine. In detail, affine misaligned was performed using random (non-isotropic) scaling, with the scaling parameter sampled from $\mathcal{U}(1 - 0.5 \cdot \alpha_a,1 + 0.5 \cdot \alpha_a)$, rotation, parameters sampled from $\mathcal{U}(-22.5 \cdot \alpha_a, 22.5 \cdot \alpha_a)$, and translation, with the parameter sampled from $\mathcal{U}(0, 0.05 \cdot \alpha_a)$ with tri-linear interpolation. Sample results of this unalignment on the CBCTLiTS dataset are shown in Fig.~\ref{Fig::DataAugResults}. All datasets were unaligned in the same way.

To quantify the unalignment effect of $\alpha_a$, the average voxel distance between the original and unaligned volumes was calculated using the perfectly paired CBCTLiTS dataset. The resulting average voxel-wise displacement (in voxels) for different $\alpha_a$ values were $\alpha_a=0 \rightarrow 0,\alpha_a=0.125 \rightarrow 2.6,\alpha_a=0.25 \rightarrow 3.9,\alpha_a=0.5 \rightarrow 6,\alpha_a=1 \rightarrow 7.8$, corresponding to physical distances of approximately $0\mathrm{mm}, 2.1\mathrm{mm}, 3.1\mathrm{mm}, 4.8\mathrm{mm}, 6.3\mathrm{mm}$, respectively. These values cover a realistic range from mild misalignment (typical of breathing-induced motion) to more substantial, clinically relevant misalignments, in line with previously reported registration errors of up to 10mm for lung SBRT patients during CBCT-CT alignment~\cite{oechsner2016registration}.

\subsection{Experimental Details}
All models were trained on a Ubuntu server using NVIDIA RTX A6000 graphics cards, using Pytorch. Due to the large data size and 48 GB VRAM memory limit, CBCTLiTS, pancreas data and synthrad volumes were isotropically downscaled by a factor of two~\cite{tschuchnig2024multi}. The dataset was split into training, validation, and testing subsets using a ratio of $0.7$ (training), $0.2$ (validation), and $0.1$ (testing). All experiments were conducted and evaluated using four different random splits to ensure stable results and to report averaged metrics and their standard deviation. To enable comparability, the same data splits and random CT misalignments were applied across all model configurations. All model variants were trained for $100$ epochs using the Adam optimizer, with a learning rate of $1 \cdot 10^{-4}$, a weight decay of $1 \cdot 10^{-5}$ for regularization, and gradient accumulation over $8$ steps (effective batch size $8$) with a per-step batch size of $1$.

\section{Results}
Tables~\ref{tab:cbct_lits_results}, \ref{tab:cbct_pancreas_results}, and \ref{tab:cbct_synthrad_results} summarize the quantitative results of sCT generation across three datasets: synthetic CBCTLiTS, pancreas, and synthrad. The data reported are mean values over four different train-val-test splits, with added standard deviations. Both unimodal and CT-only baselines are given. Overall, the proposed method improves on the baseline multimodal method in $79$ out of $90$ experimental setups. On the unimodal baseline, the STM+MM method achieves improved results in $83$ out of $90$ cases and against the CT-only baseline in $6$ out of $72$ cases. Overall, CT-only baselines show little difference between MM+STN and MM.

On the CBCTLiTS dataset, our proposed MM+STN model consistently outperforms both the multimodal (MM) setup without STN as well as the unimodal and CT-only baselines. The addition, MM+STN has the biggest positive effects on low to moderately aligned setups with low intraoperative CBCT visual quality. In the pancreas dataset, MM+STN improves over MM in all but two cases. Compared to the unimodal and CT-only baselines applied to the pancreas dataset, STM+MM improve results in $12$ out of $15$ (unimodal) and $8$ out of $12$ (CT-only) setups respectively. The synthrad dataset shows inconclusive results. While MAE and SSIM metrics worsen in most ($70\%$) cases, Perceptual metrics improve in four out of five cases. In detail, compared to MM, MM+STN improves results in seven out of $15$ cases, $11$ out of $15$ cases compared to the unimodal baseline and $10$ out of $12$ cases compared to CT-only.

\begin{table}[t]
\centering
\caption{Evaluation results (mean $\pm$ std) on the Pancreas Dataset. The results shown include unimodal baselines (Base), multimodal model results~\cite{tschuchnig2025enhancing} (MM) and the proposed MM+STN model results. Values in bold show improvements towards all baselines.}
\label{tab:cbct_pancreas_results}
\resizebox{\textwidth}{!}{%
\begin{tabular}{c | c | c | c | c | c | c | c}
\hline
& $\alpha_a$ & MAE & CT-only & \(\mathrm{1\text{-}SSIM}\) & CT-only & Perceptual & CT-only \\
\hline
Base & & 0.111 $\pm$ 0.037 & & 0.094 $\pm$ 0.026 & & 0.225 $\pm$ 0.025 & \\ \hline
\multirow{5}{*}{MM} & 1 & 0.114 $\pm$ 0.038 & 0.243 $\pm$ 0.070 & 0.108 $\pm$ 0.020 & 0.169 $\pm$ 0.049 & 0.227 $\pm$ 0.028 & 0.350 $\pm$ 0.065 \\
& 0.5 & 0.114 $\pm$ 0.039 & 0.150 $\pm$ 0.044 & 0.105 $\pm$ 0.018 & 0.135 $\pm$ 0.038 & 0.232 $\pm$ 0.024 & 0.272 $\pm$ 0.054 \\
& 0.25 & 0.105 $\pm$ 0.038 & 0.088 $\pm$ 0.027 & 0.100 $\pm$ 0.027 & 0.101 $\pm$ 0.028 & 0.213 $\pm$ 0.022 & 0.195 $\pm$ 0.038 \\
& 0.125 & 0.087 $\pm$ 0.033 & 0.051 $\pm$ 0.016 & 0.091 $\pm$ 0.020 & 0.068 $\pm$ 0.022 & 0.172 $\pm$ 0.020 & 0.136 $\pm$ 0.027 \\
& 0 & 0.045 $\pm$ 0.027 & 0.000 $\pm$ 0.000 & 0.031 $\pm$ 0.013 & 0.000 $\pm$ 0.000 & 0.096 $\pm$ 0.012 & 0.000 $\pm$ 0.000 \\
\hline
\multirow{5}{*}{MM$+$STN} & 1 & \textbf{0.110 $\pm$ 0.044} & 0.246 $\pm$ 0.060 & 0.124 $\pm$ 0.058 & 0.164 $\pm$ 0.048 & 0.228 $\pm$ 0.023 & 0.359 $\pm$ 0.042 \\
& 0.5 & \textbf{0.107 $\pm$ 0.036} & 0.155 $\pm$ 0.029 & \textbf{0.104 $\pm$ 0.032} & 0.134 $\pm$ 0.030 & \textbf{0.221 $\pm$ 0.025} & 0.289 $\pm$ 0.029 \\
& 0.25 & \textbf{0.094 $\pm$ 0.028} & 0.093 $\pm$ 0.015 & \textbf{0.090 $\pm$ 0.018} & 0.104 $\pm$ 0.020 & \textbf{0.205 $\pm$ 0.023} & 0.206 $\pm$ 0.023 \\
& 0.125 & \textbf{0.080 $\pm$ 0.027} & 0.053 $\pm$ 0.009 & \textbf{0.071 $\pm$ 0.019} & 0.071 $\pm$ 0.015 & \textbf{0.155 $\pm$ 0.015} & 0.142 $\pm$ 0.016 \\
& 0 & \textbf{0.041 $\pm$ 0.027} & 0.000 $\pm$ 0.000 & \textbf{0.018 $\pm$ 0.010} & 0.000 $\pm$ 0.000 & \textbf{0.082 $\pm$ 0.005} & 0.000 $\pm$ 0.000 \\
\hline
\end{tabular}
}
\end{table}

\begin{table}[t]
\centering
\caption{Evaluation results (mean $\pm$ std) on the SynthRad Dataset. The results shown include unimodal baselines (Base), multimodal model results~\cite{tschuchnig2025enhancing} (MM) and the proposed MM+STN model results. Values in bold show improvements towards all baselines.}
\label{tab:cbct_synthrad_results}
\resizebox{\textwidth}{!}{%
\begin{tabular}{c | c | c | c | c | c | c | c}
\hline
& $\alpha_a$ & MAE & CT-only & \(\mathrm{1\text{-}SSIM}\) & CT-only & Perceptual & CT-only \\
\hline
Base & & 0.127 $\pm$ 0.059 & & 0.149 $\pm$ 0.138 & & 0.188 $\pm$ 0.039 & \\ \hline
\multirow{5}{*}{MM} & 1 & 0.120 $\pm$ 0.067 & 0.304 $\pm$ 0.091 & 0.136 $\pm$ 0.107 & 0.285 $\pm$ 0.110 & 0.198 $\pm$ 0.038 & 0.420 $\pm$ 0.068 \\
 & 0.5 & 0.121 $\pm$ 0.061 & 0.187 $\pm$ 0.053 & 0.119 $\pm$ 0.102 & 0.194 $\pm$ 0.071 & 0.196 $\pm$ 0.038 & 0.350 $\pm$ 0.056 \\
& 0.25 & 0.111 $\pm$ 0.050 & 0.114 $\pm$ 0.032 & 0.131 $\pm$ 0.116 & 0.131 $\pm$ 0.046 & 0.196 $\pm$ 0.035 & 0.257 $\pm$ 0.044 \\
& 0.125 & 0.101 $\pm$ 0.047 & 0.069 $\pm$ 0.020 & 0.128 $\pm$ 0.109 & 0.086 $\pm$ 0.032 & 0.184 $\pm$ 0.028 & 0.181 $\pm$ 0.034 \\
& 0 & 0.055 $\pm$ 0.028 & 0.000 $\pm$ 0.000 & 0.077 $\pm$ 0.111 & 0.000 $\pm$ 0.000 & 0.101 $\pm$ 0.021 & 0.000 $\pm$ 0.000 \\
\hline
\multirow{5}{*}{MM$+$STN} & 1 & 0.127 $\pm$ 0.063 & 0.312 $\pm$ 0.083 & 0.142 $\pm$ 0.122 & 0.289 $\pm$ 0.109 & 0.198 $\pm$ 0.037 & 0.429 $\pm$ 0.058 \\
& 0.5 & 0.121 $\pm$ 0.055 & 0.194 $\pm$ 0.047 & 0.146 $\pm$ 0.122 & 0.198 $\pm$ 0.071 & \textbf{0.191 $\pm$ 0.037} & 0.358 $\pm$ 0.049 \\
& 0.25 & 0.114 $\pm$ 0.054 & 0.117 $\pm$ 0.029 & \textbf{0.123 $\pm$ 0.103} & 0.135 $\pm$ 0.048 & \textbf{0.192 $\pm$ 0.037} & 0.264 $\pm$ 0.042 \\
& 0.125 & \textbf{0.097 $\pm$ 0.043} & 0.071 $\pm$ 0.018 & 0.130 $\pm$ 0.114 & 0.089 $\pm$ 0.034 & \textbf{0.163 $\pm$ 0.024} & 0.186 $\pm$ 0.032 \\
& 0 & \textbf{0.053 $\pm$ 0.025} & 0.000 $\pm$ 0.000 & 0.077 $\pm$ 0.109 & 0.000 $\pm$ 0.000 & \textbf{0.096 $\pm$ 0.018} & 0.000 $\pm$ 0.000 \\
\hline
\end{tabular}
}
\end{table}

\begin{table}[t]
\centering
\caption{Evaluation results (mean $\pm$ std) on the CBCTLiTS Dataset. The results shown include unimodal baselines (Base), multimodal model results~\cite{tschuchnig2025enhancing} (MM) and the proposed MM+STN model results. Values in bold show improvements towards all baselines.}
\label{tab:cbct_lits_results}
\resizebox{\textwidth}{!}{%
\begin{tabular}{c | c | c | c | c | c | c | c | c}
\hline
& $\alpha_{np}$ & $\alpha_a$ & MAE & CT-only & \(\mathrm{1\text{-}SSIM}\) & CT-only & Perceptual & CT-only \\
\hline
Base & 32 & & 0.306 $\pm$ 0.128 & & 0.399 $\pm$ 0.064 & & 0.391 $\pm$ 0.056 & \\ \hline
\multirow{5}{*}{MM} & 32 & 1 & 0.302 $\pm$ 0.116 & 0.764 $\pm$ 0.216 & 0.396 $\pm$ 0.070 & 0.668 $\pm$ 0.097 & 0.392 $\pm$ 0.054 & 0.666 $\pm$ 0.056 \\
& 32 & 0.5 & 0.280 $\pm$ 0.101 & 0.525 $\pm$ 0.124 & 0.361 $\pm$ 0.066 & 0.561 $\pm$ 0.079 & 0.388 $\pm$ 0.055 & 0.594 $\pm$ 0.061 \\
& 32 & 0.25 & 0.241 $\pm$ 0.075 & 0.348 $\pm$ 0.076 & 0.354 $\pm$ 0.075 & 0.458 $\pm$ 0.074 & 0.380 $\pm$ 0.054 & 0.475 $\pm$ 0.063 \\
& 32 & 0.125 & 0.206 $\pm$ 0.057 & 0.219 $\pm$ 0.050 & 0.320 $\pm$ 0.066 & 0.341 $\pm$ 0.066 & 0.353 $\pm$ 0.055 & 0.343 $\pm$ 0.057 \\
& 32 & 0 & 0.060 $\pm$ 0.020 & 0.000 $\pm$ 0.000 & 0.086 $\pm$ 0.044 & 0.000 $\pm$ 0.000 & 0.187 $\pm$ 0.019 & 0.000 $\pm$ 0.000 \\
\hline

\multirow{5}{*}{MM$+$STN} & 32 & 1 & \textbf{0.281 $\pm$ 0.114} & 0.847 $\pm$ 0.267 & \textbf{0.357 $\pm$ 0.070} & 0.696 $\pm$ 0.104 & \textbf{0.380 $\pm$ 0.055} & 0.670 $\pm$ 0.081 \\
& 32 & 0.5 & \textbf{0.253 $\pm$ 0.095} & 0.575 $\pm$ 0.158 & \textbf{0.323 $\pm$ 0.062} & 0.582 $\pm$ 0.086 & \textbf{0.378 $\pm$ 0.055} & 0.617 $\pm$ 0.082 \\
& 32 & 0.25 & \textbf{0.207 $\pm$ 0.068} & 0.380 $\pm$ 0.092 & \textbf{0.303 $\pm$ 0.088} & 0.477 $\pm$ 0.077 & \textbf{0.348 $\pm$ 0.059} & 0.504 $\pm$ 0.079 \\
& 32 & 0.125 & \textbf{0.164 $\pm$ 0.049} & 0.241 $\pm$ 0.055 & \textbf{0.234 $\pm$ 0.086} & 0.363 $\pm$ 0.069 & \textbf{0.316 $\pm$ 0.050} & 0.372 $\pm$ 0.071 \\
& 32 & 0 & 0.062 $\pm$ 0.019 & 0.000 $\pm$ 0.000 & \textbf{0.081 $\pm$ 0.042} & 0.000 $\pm$ 0.000 & \textbf{0.185 $\pm$ 0.018} & 0.000 $\pm$ 0.000 \\

\hline

Base & 64 & & 0.289 $\pm$ 0.122 & & 0.358 $\pm$ 0.061 & & 0.380 $\pm$ 0.057 & \\ \hline
\multirow{5}{*}{MM} & 64 & 1 & 0.282 $\pm$ 0.112 & 0.764 $\pm$ 0.216 & 0.357 $\pm$ 0.060 & 0.668 $\pm$ 0.097 & 0.380 $\pm$ 0.055 & 0.666 $\pm$ 0.056 \\
& 64 & 0.5 & 0.262 $\pm$ 0.093 & 0.525 $\pm$ 0.124 & 0.326 $\pm$ 0.057 & 0.561 $\pm$ 0.079 & 0.377 $\pm$ 0.056 & 0.594 $\pm$ 0.061 \\
& 64 & 0.25 & 0.233 $\pm$ 0.082 & 0.348 $\pm$ 0.076 & 0.316 $\pm$ 0.074 & 0.458 $\pm$ 0.074 & 0.371 $\pm$ 0.056 & 0.475 $\pm$ 0.063 \\
& 64 & 0.125 & 0.200 $\pm$ 0.062 & 0.219 $\pm$ 0.050 & 0.316 $\pm$ 0.073 & 0.341 $\pm$ 0.066 & 0.343 $\pm$ 0.054 & 0.343 $\pm$ 0.057 \\
& 64 & 0 & 0.069 $\pm$ 0.016 & 0.000 $\pm$ 0.000 & 0.096 $\pm$ 0.049 & 0.000 $\pm$ 0.000 & 0.194 $\pm$ 0.017 & 0.000 $\pm$ 0.000 \\
\hline

\multirow{5}{*}{MM$+$STN} & 64 & 1 & \textbf{0.268 $\pm$ 0.118} & 0.847 $\pm$ 0.267 & \textbf{0.322 $\pm$ 0.064} & 0.696 $\pm$ 0.104 & \textbf{0.370 $\pm$ 0.057} & 0.670 $\pm$ 0.081 \\
& 64 & 0.5 & \textbf{0.243 $\pm$ 0.093} & 0.575 $\pm$ 0.158 & \textbf{0.300 $\pm$ 0.067} & 0.582 $\pm$ 0.086 & \textbf{0.365 $\pm$ 0.055} & 0.617 $\pm$ 0.082 \\
& 64 & 0.25 & \textbf{0.200 $\pm$ 0.069} & 0.380 $\pm$ 0.092 & \textbf{0.276 $\pm$ 0.074} & 0.477 $\pm$ 0.077 & \textbf{0.346 $\pm$ 0.054} & 0.504 $\pm$ 0.079 \\
& 64 & 0.125 & \textbf{0.161 $\pm$ 0.049} & 0.241 $\pm$ 0.055 & \textbf{0.227 $\pm$ 0.082} & 0.363 $\pm$ 0.069 & \textbf{0.313 $\pm$ 0.052} & 0.372 $\pm$ 0.071 \\
& 64 & 0 & \textbf{0.062 $\pm$ 0.015} & 0.000 $\pm$ 0.000 & \textbf{0.087 $\pm$ 0.048} & 0.000 $\pm$ 0.000 & \textbf{0.185 $\pm$ 0.016} & 0.000 $\pm$ 0.000 \\

\hline

Base & 128 & & 0.281 $\pm$ 0.122 & & 0.328 $\pm$ 0.068 & & 0.372 $\pm$ 0.059 & \\ \hline
\multirow{5}{*}{MM} & 128 & 1 & 0.275 $\pm$ 0.113 & 0.764 $\pm$ 0.216 & 0.331 $\pm$ 0.072 & 0.668 $\pm$ 0.097 & 0.369 $\pm$ 0.055 & 0.666 $\pm$ 0.056 \\
& 128 & 0.5 & 0.250 $\pm$ 0.102 & 0.525 $\pm$ 0.124 & 0.296 $\pm$ 0.070 & 0.561 $\pm$ 0.079 & 0.368 $\pm$ 0.057 & 0.594 $\pm$ 0.061 \\
& 128 & 0.25 & 0.219 $\pm$ 0.087 & 0.348 $\pm$ 0.076 & 0.293 $\pm$ 0.069 & 0.458 $\pm$ 0.074 & 0.362 $\pm$ 0.055 & 0.475 $\pm$ 0.063 \\
& 128 & 0.125 & 0.191 $\pm$ 0.063 & 0.219 $\pm$ 0.050 & 0.280 $\pm$ 0.076 & 0.341 $\pm$ 0.066 & 0.343 $\pm$ 0.051 & 0.343 $\pm$ 0.057 \\
& 128 & 0 & 0.074 $\pm$ 0.021 & 0.000 $\pm$ 0.000 & 0.097 $\pm$ 0.048 & 0.000 $\pm$ 0.000 & 0.192 $\pm$ 0.016 & 0.000 $\pm$ 0.000 \\
\hline

\multirow{5}{*}{MM$+$STN} & 128 & 1 & \textbf{0.263 $\pm$ 0.104} & 0.847 $\pm$ 0.267 & \textbf{0.301 $\pm$ 0.073} & 0.696 $\pm$ 0.104 & \textbf{0.361 $\pm$ 0.053} & 0.670 $\pm$ 0.081 \\
& 128 & 0.5 & \textbf{0.242 $\pm$ 0.103} & 0.575 $\pm$ 0.158 & \textbf{0.267 $\pm$ 0.079} & 0.582 $\pm$ 0.086 & \textbf{0.361 $\pm$ 0.058} & 0.617 $\pm$ 0.082 \\
& 128 & 0.25 & \textbf{0.203 $\pm$ 0.079} & 0.380 $\pm$ 0.092 & \textbf{0.249 $\pm$ 0.077} & 0.477 $\pm$ 0.077 & \textbf{0.349 $\pm$ 0.056} & 0.504 $\pm$ 0.079 \\
& 128 & 0.125 & \textbf{0.160 $\pm$ 0.051} & 0.241 $\pm$ 0.055 & \textbf{0.211 $\pm$ 0.081} & 0.363 $\pm$ 0.069 & \textbf{0.309 $\pm$ 0.054} & 0.372 $\pm$ 0.071 \\
& 128 & 0 & \textbf{0.065 $\pm$ 0.021} & 0.000 $\pm$ 0.000 & \textbf{0.094 $\pm$ 0.054} & 0.000 $\pm$ 0.000 & \textbf{0.188 $\pm$ 0.014} & 0.000 $\pm$ 0.000 \\

\hline

Base & 256 & & 0.275 $\pm$ 0.126 & & 0.313 $\pm$ 0.072 & & 0.360 $\pm$ 0.062 & \\ \hline
\multirow{5}{*}{MM} & 256 & 1 & 0.273 $\pm$ 0.118 & 0.764 $\pm$ 0.216 & 0.312 $\pm$ 0.068 & 0.668 $\pm$ 0.097 & 0.363 $\pm$ 0.060 & 0.666 $\pm$ 0.056 \\
& 256 & 0.5 & 0.250 $\pm$ 0.097 & 0.525 $\pm$ 0.124 & 0.268 $\pm$ 0.069 & 0.561 $\pm$ 0.079 & 0.357 $\pm$ 0.059 & 0.594 $\pm$ 0.061 \\
& 256 & 0.25 & 0.222 $\pm$ 0.087 & 0.348 $\pm$ 0.076 & 0.264 $\pm$ 0.068 & 0.458 $\pm$ 0.074 & 0.353 $\pm$ 0.057 & 0.475 $\pm$ 0.063 \\
& 256 & 0.125 & 0.184 $\pm$ 0.063 & 0.219 $\pm$ 0.050 & 0.260 $\pm$ 0.070 & 0.341 $\pm$ 0.066 & 0.339 $\pm$ 0.050 & 0.343 $\pm$ 0.057 \\
& 256 & 0 & 0.070 $\pm$ 0.023 & 0.000 $\pm$ 0.000 & 0.095 $\pm$ 0.049 & 0.000 $\pm$ 0.000 & 0.197 $\pm$ 0.014 & 0.000 $\pm$ 0.000 \\
\hline

\multirow{5}{*}{MM$+$STN} & 256 & 1 & \textbf{0.267 $\pm$ 0.116} & 0.847 $\pm$ 0.267 & \textbf{0.295 $\pm$ 0.074} & 0.696 $\pm$ 0.104 & \textbf{0.354 $\pm$ 0.058} & 0.670 $\pm$ 0.081 \\
& 256 & 0.5 & \textbf{0.239 $\pm$ 0.100} & 0.575 $\pm$ 0.158 & \textbf{0.259 $\pm$ 0.076} & 0.582 $\pm$ 0.086 & \textbf{0.350 $\pm$ 0.058} & 0.617 $\pm$ 0.082 \\
& 256 & 0.25 & \textbf{0.203 $\pm$ 0.073} & 0.380 $\pm$ 0.092 & \textbf{0.231 $\pm$ 0.070} & 0.477 $\pm$ 0.077 & \textbf{0.338 $\pm$ 0.058} & 0.504 $\pm$ 0.079 \\
& 256 & 0.125 & \textbf{0.151 $\pm$ 0.046} & 0.241 $\pm$ 0.055 & \textbf{0.198 $\pm$ 0.072} & 0.363 $\pm$ 0.069 & \textbf{0.303 $\pm$ 0.049} & 0.372 $\pm$ 0.071 \\
& 256 & 0 & \textbf{0.065 $\pm$ 0.016} & 0.000 $\pm$ 0.000 & \textbf{0.091 $\pm$ 0.051} & 0.000 $\pm$ 0.000 & \textbf{0.188 $\pm$ 0.013} & 0.000 $\pm$ 0.000 \\

\hline
\end{tabular}
}
\end{table}

\begin{figure}[t]
\includegraphics[width=\textwidth]{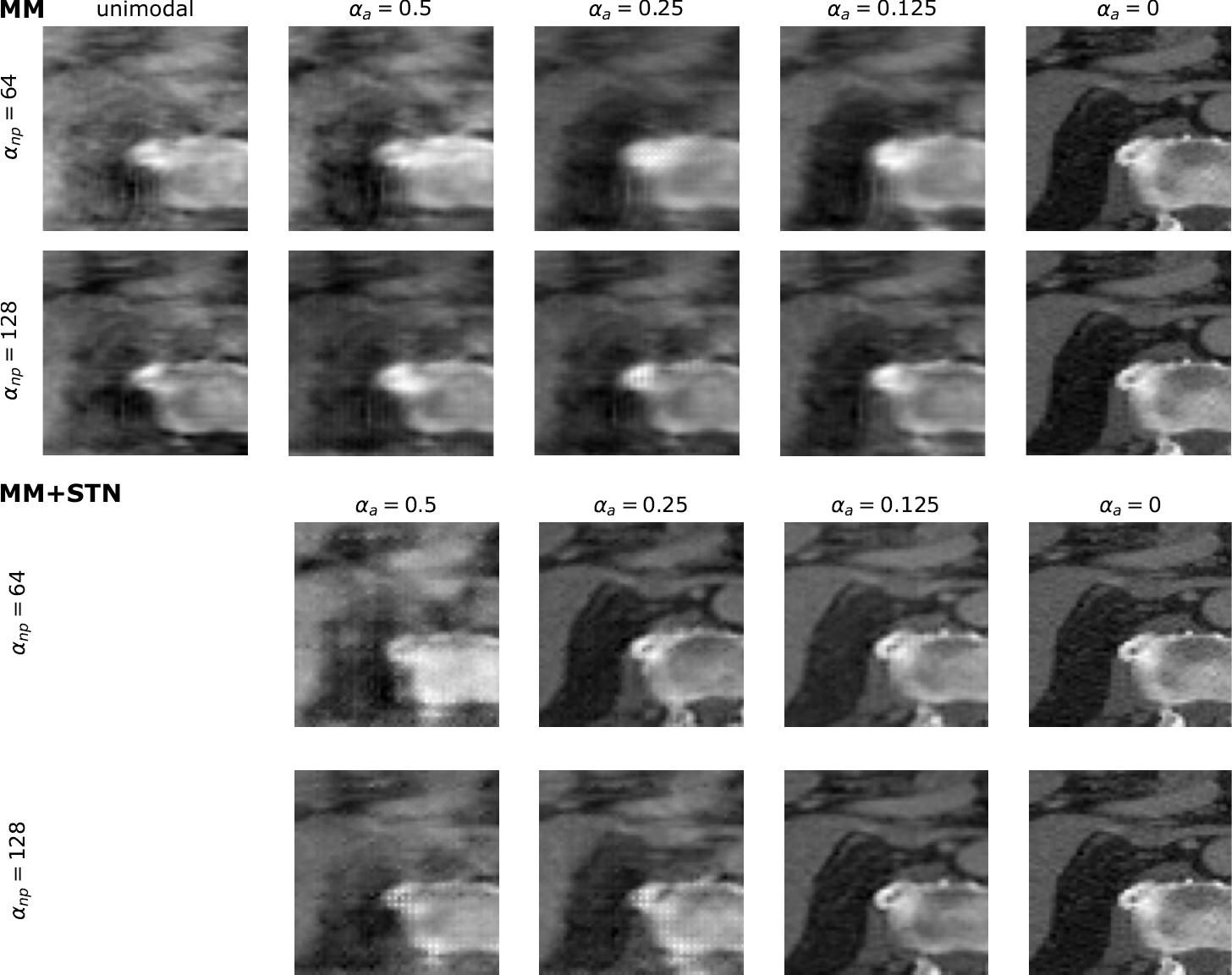}
\caption{
Qualitative comparison of sCT reconstructions on the CBCTLiTS dataset. Rows correspond to different CBCT quality levels (top: low, bottom: high). Columns compare methods: a unimodal CBCT baseline (left) and multimodal reconstructions conditioned on preoperative CTs with increasing alignment (right). Images are normalized for visualization and shown at increased zoom to better highlight anatomical structures. Results of both multimodal (MM) and multimodal with STN registration (MM+STN) are shown.}
\label{Fig::qual}
\end{figure}

Fig.~\ref{Fig::qual} qualitatively evaluates a randomly sampled sCT, generated from different source quality CBCT, CBCT-CT alignment and sCT generation method (multimodal as MM and multimodal with registration as MM+STN).

\section{Discussion}
Our results on synthetic data demonstrate that the addition of STN into the MM model systematically improves alignment between preoperative CT and intraoperative CBCT, generally enhancing sCT generation ($79$ out of $90$ cases). In nearly all setups with moderate to low misalignment, the inclusion of STN led to consistent improvements over the multimodal model without STN, the unimodal CBCT baseline, and the CT-only baseline. This confirms that end-to-end learning with STN is effective for compensating moderate anatomical discrepancies, particularly when training data closely resemble the synthetic setup where near-perfect CBCT-CT alignment exists.

However, performance declines at high misalignment levels (\(\alpha_a = 1\)), suggesting that STN struggles with extreme spatial transformations. This is especially noticable in real-world data scenarios with fuzzy ground truth registration targets. In such cases, external pre-registration, possibly informed by clinical expertise, may be necessary to bring inputs into a roughly aligned space before fine-tuning with STN. Also, for extreme misalignment, reverting to unimodal sCT models may be more effective due to their simplicity and consistency.

Results on real-world clinical datasets largely mirror trends from synthetic data, indicating strong generalizability. While the SynthRad dataset yielded mixed results (improvements in $47\%$ of cases), the pancreas dataset showed notable improvements (improvements in $87\%$ of cases) over the multimodal method without STN. Interestingly, STN also provided benefits in well-aligned settings, potentially due to its ability to capture subtle anatomical variations not addressed by conventional registration or static model assumptions present in the used real-world datasets.

Despite its benefits, STN occasionally led to decreases in MAE and \(\mathrm{1\text{-}SSIM}\) on the SynthRad dataset. However, perceptual metrics generally improved across all datasets and setups. This suggests that STN may enhance spatial coherence and anatomical realism even if pixelwise similarity does not improve uniformly. This qualitative improvement is especially visible in the presented CBCTLiTS qualitative results figure, where STN clearly enhances fine structural detail.

To conclude, we demonstrated how the STN can be seamlessly integrated into a U-Net based sCT generation framework to address anatomical misalignment in multimodal imaging and how this additional STN reacts to the parameters of intraoperative CBCT quality and CBCT-CT alignment. Our findings suggest that STN-based alignment consistently enhances sCT quality across both synthetic and clinical datasets, particularly under moderate misalignment. We find STN to be especially useful if the intraoperative volume is of low visual quality. These insights highlight the potential of learnable registration to improve intraoperative imaging workflows and motivate further research into robust, adaptive multimodal learning strategies for clinical deployment.
\begin{credits}
\subsubsection{\ackname} This project was partly funded by the Austrian Research Promotion Agency (FFG) under the bridge project "CIRCUIT: Towards Comprehensive CBCT Imaging Pipelines for Real-time Acquisition, Analysis, Interaction and Visualization" (CIRCUIT), no. 41545455 and by the county of Salzburg under the project AIBIA and the Salzburg University of Applied Sciences under the project FHS-trampoline-8 (Applied Data Science Lab).

\subsubsection{\discintname}
%It is now necessary to declare any competing interests or to specifically
%state that the authors have no competing interests. Please place the
%statement with a bold run-in heading in small font size beneath the
%(optional) acknowledgments\footnote{If EquinOCS, our proceedings submission system, is used, then the disclaimer can be provided directly in the system.},
%for example: The authors have no competing interests to declare that are
%relevant to the content of this article. Or: Author A has received research
%grants from Company W. Author B has received a speaker honorarium from
%Company X and owns stock in Company Y. Author C is a member of committee Z.
Not applicable
\end{credits}
%
% ---- Bibliography ----
%
% BibTeX users should specify bibliography style 'splncs04'.
% References will then be sorted and formatted in the correct style.
%
\bibliographystyle{splncs04}
\bibliography{mybibliography}
\end{document}